\documentclass[twoside]{article}
\usepackage{fleqn,espcrc2}

\usepackage{graphicx}

\newcommand{\AmS}{{\protect\the\textfont2
  A\kern-.1667em\lower.5ex\hbox{M}\kern-.125emS}}

\title{Systematic evolution of the magnetotransport properties of 
Bi$_{2}$Sr$_{2-x}$La$_x$CuO$_{6}$ in a wide doping range}

\author{Yoichi Ando,\address{Central Research Institute of 
Electric Power Industry (CRIEPI), Komae, Tokyo 201-8511, 
Japan}\address{Department of Physics, Science University of Tokyo, 
Shinjuku-ku, Tokyo 162-8601, Japan}
T. Murayama,$^{\rm{ab}}$\thanks{Present address: Nichia Corporation,
Kaminaka-cho, Anan-shi, Tokushima 774-8601, Japan.}
and S. Ono$^{\rm{a}}$}
       
\begin{document}

\begin{abstract}
Recently we have succeeded in growing 
a series of high-quality Bi$_{2}$Sr$_{2-x}$La$_x$CuO$_{6}$ 
crystals in a wide range of carrier concentrations.  
The data of $\rho_{ab}(T)$ and $R_H(T)$ of those crystals
show behaviors that are considered to be ``canonical"
to the cuprates.
The optimum zero-resistance $T_c$ has been raised to 
as high as 38 K, which is almost equal to the 
optimum $T_c$ of La$_{2-x}$Sr$_{x}$CuO$_{4}$.
\vspace{1pc}
\end{abstract}

\maketitle

\section{INTRODUCTION}

Since the high-$T_c$ cuprates are in essence doped Mott insulators,
systematic studies of the evolution of the normal-state 
properties upon changing the carrier concentration are 
very useful for elucidating the origin of the peculiar normal state.
Bi$_{2}$Sr$_{2}$CuO$_{6}$ (Bi-2201) system is an attractive 
candidate for such studies, because the carrier concentration can be 
widely changed by partially replacing Sr with La \cite{Maeda}.  
Moreover, this system allows us 
to study the normal-state in a wider temperature range, because 
the optimum $T_c$ (achieved in Bi$_{2}$Sr$_{2-x}$La$_x$CuO$_{6}$ with 
$x$$\simeq$0.4 \cite{Maeda,Yoshizaki}) has been reported to be about 
30 K, which is lower than the optimum $T_c$ of 
La$_{2-x}$Sr$_{x}$CuO$_{4}$ (LSCO). 
However, a number of problems have been known so far for
Bi-2201 crystals: 
(i) the transport properties of Bi-2201
are quite non-reproducible even among crystals of nominally the 
same composition \cite{Mackenzie,Ando};
(ii) the residual resistivity of $\rho_{ab}$ 
is larger (the smallest value reported to date is 
70 $\mu \Omega$cm \cite{Ando,Martin}) than other cuprates; and 
(iii) the temperature dependence of the Hall coefficient $R_H$ is weak 
and thus the cotangent of the Hall angle $\theta_H$ does not obey the 
$T^2$ law \cite{Mackenzie}.

In our group at CRIEPI, we have recently succeeded in growing
a series of high-quality crystals, in which 
the above problems have mostly been overcome \cite{Murayama}.  
Here we report most recent data of $\rho_{ab}(T)$ and $R_H(T)$ of 
our Bi-2201 crystals in a wide range
of carrier concentrations to demonstrate that 
the normal-state transport properties in 
those clean crystals display behaviors
that are in good accord with other cuprates.

\section{SAMPLES}

The single crystals of Bi$_{2}$Sr$_{2-x}$La$_x$CuO$_{6}$ (BSLCO) 
are grown using a floating-zone technique.  
The crystals are annealed in oxygen
to sharpen the superconducting transition width.
Pure Bi-2201 is an overdoped system \cite{Maeda} and 
increasing La doping brings the system from overdoped region 
to underdoped region.
In our series of crystals, 
the optimum doping is achieved with $x$$\simeq$0.4, which is consistent 
with previous reports on BSLCO \cite{Maeda,Yoshizaki}.
The actual La concentrations in the crystals are 
determined with the inductively-coupled plasma (ICP) analysis.
We note that the optimum zero-resistance $T_c$ reported here is
as high as 38 K, which is not only the highest
value ever reported for Bi-2201 system but also almost equals
that of the LSCO system.

\begin{figure}[!b!]
\vspace{-10mm}
\centerline{\includegraphics[scale=0.40]{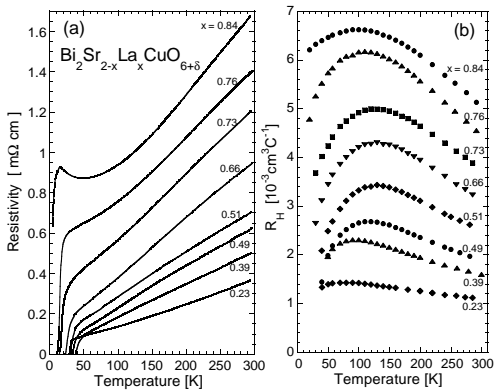}}
\vspace{-57mm}
\caption{$T$ dependence of (a) $\rho_{ab}$ and (b) $R_H$ of the 
BSLCO crystals with various $x$. Note that the extrapolated
residual resistivity of $x$=0.39 sample is 25 $\mu \Omega$cm, which
is the smallest value to date for Bi-2201 or BSLCO.}  
\label{fig1} 
\end{figure}

\section{RESULTS AND DISCUSSIONS}

Figure 1(a) shows the $T$ dependence of $\rho_{ab}$ for 
eight $x$ values ($x$=0.23, 0.39, 0.49, 0.51, 0.66, 0.73, 0.76 and 0.84) 
in zero field.
Clearly, both the magnitude of $\rho_{ab}$ and its slope  
show systematic decrease with increasing carrier concentration
(decreasing $x$).
Note that it is only at the optimum doping ($x$=0.39) that $\rho_{ab}$
shows a good $T$-linear behavior: In the underdoped region, 
$\rho_{ab}(T)$ shows a downward deviation from the $T$-linear 
behavior, which has been discussed to mark the pseudogap
\cite{Ito}. In the
overdoped region, $\rho_{ab}(T)$ shows an upward curvature in the whole
temperature range; the $T$ dependence of $\rho_{ab}$ in the
overdoped region can be well described by $\rho_{ab}$=$\rho_0+AT^n$ 
(with $n$$\approx$1.2 for $x$=0.23),
which is a behavior known to be peculiar for the overdoped cuprates
\cite{Kubo,Takagi}.

Shown in Fig. 1(b) is the $T$ dependence of $R_H$ for the eight 
samples.  Here again, a clear evolution of $R_H$ with $x$ is observed; 
the change in the magnitude of $R_H$ at 300 K suggests that the carrier 
concentration is actually reduced roughly by a factor of 5 upon 
increasing $x$ from 0.23 to 0.84.  Note that the $T$ dependence of
$R_H$ is stronger than those previously reported 
\cite{Mackenzie,Ando2} and that pronounced peaks in $R_H(T)$ 
are clearly observed in optimally-doped and underdoped samples.

When $\cot \theta_H$ is examined, we found that $\cot \theta_H$ obeys
a power-law dependence $T^{\alpha}$, where $\alpha$ is
nearly 2 in underdoped samples 
but shows a systematic decrease with increasing carrier concentration
\cite{Murayama}.
Figure 2 shows examples of the $\cot \theta_H$ vs 
$T^{\alpha}$ plot, for $x$=0.66 and 0.39.  
We note that the $T^2$ law of $\cot \theta_H$
is confirmed for the first time for Bi-2201 in our crystals.

\begin{figure}[bt]
\vspace{-27mm}
\centerline{\includegraphics[scale=0.46]{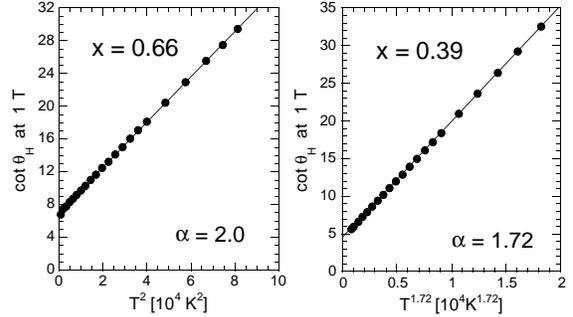}}
\vspace{-69mm}
\caption{$\cot \theta_H$ vs $T^{\alpha}$ plot for 
$x$=0.66 and 0.39.}  
\label{fig2} 
\end{figure}

\section{CONCLUSION}

We present the $\rho_{ab}(T)$ and $R_H(T)$ data of a series of high-quality 
La-doped Bi-2201 crystals in a wide range of carrier concentrations.
It is shown that the optimum zero-resistance $T_c$ of Bi-2201 can be as high
as 38 K.
The normal-state transport properties of our Bi-2201 crystals
show systematics that are in good accord with other cuprates.

\end{document}